\documentstyle[11pt,twoside,sympo2011,epsfig]{article}
\begin{document}
\title{Investigating stellar activity with CoRoT data and complementary ground-based observations
}
\author{S. Mathur$^1$, D. Salabert$^2$, R. A. Garc\'\i a$^3$, T. S. Metcalfe$^1$, C. R\'egulo$^{4,5}$, J. Ballot$^6$ 
}
\affil{
$^1$High Altitude Observatory, NCAR, P.O. Box 3000, Boulder, CO 80307, USA\\
$^2$Universit\'e de Nice Sophia-Antipolis, CNRS, Observatoire de la C\^ote dÕAzur, BP 4229, 06304 Nice Cedex 4, France\\
$^3$Laboratoire AIM, CEA/DSM -- CNRS - Universit\'e Paris Diderot -- IRFU/SAp, Centre de Saclay, 91191 Gif-sur-Yvette Cedex, France\\
$^4$Universidad de La Laguna, Dpto de Astrof\'isica, 38206, Tenerife, Spain\\
$^5$Instituto de Astrof\'\i sica de Canarias, 38205, La Laguna, Tenerife, Spain\\
$^6$Institut de Recherche en Astrophysique et Plan\'etologie, Universit\'e de Toulouse, CNRS, F-31400, Toulouse, France
} 
%INAF - Osservatorio Astronomico di Brera, via E. Bianchi 46, 23807, Merate (LC), Italy
%
\begin{abstract}
Recently, the study of the CoRoT target, HD49933, showed evidence of variability of its magnetic activity. This was the first time that a stellar activity was detected using asteroseismic data. For the Sun and HD49933, we observe an increase of the p-mode frequencies and a decrease of the maximum amplitude per radial mode when the activity level is higher. Moreover, we have been able to determine the variation of the frequency shift as a function of frequency, a premier in a star different from the Sun, showing pretty interesting similarities with the behavior already seen in the Sun. Beside, surface activity is now confirmed by the continuous monitoring done in Ca H\&K during 6 months last fall. The peak-to-peak activity level seems to be bigger than 20\% (close to the solar level, which is around 25\%).
We studied some other CoRoT solar-like targets as well for which modes have been detected and well identified, e.g. HD181420 and HD52265 (which is hosting a planet). Although HD52265 rotates much slower  than HD49933 meaning much longer activity cycle (compared to the CoRoT observations), we do indeed see a hint of an increase in the activity level of that star...
\end{abstract}
\section{Introduction}
Better understanding of dynamo process is crucial for our knowledge of the interaction between rotation, convection and magnetic fields and thus to be able to better predict the activity cycle on the Sun, which is still far for being under control as seen during the unexpected long solar minima between cycles 23 and 24 (e.g. Salabert et al. 2009). Moreover, the activity of stars are an important factor to allow the development  of life based on the Carbon chemistry inside the habitable zone. It is still not clear how the planet can also have an impact on the activity of the host star (e.g. Poppenhaeger \& Schmidt, 2011).

Nowadays the discovery of short magnetic activity cycles such as the one found in i~Hor (Metcalfe et al. 2010) or HD49933 (Garc\'\i a et al. 2010), combined with the long time series provided by CoRoT (Convection, Rotation and planetary Targets, Michel et al. 2008) and {\it Kepler} (Borucki et al. 2010) of solar-like stars (e.g. Barban et al. 2009; Deheuvels et al. 2010; Garc\'\i a et al. 2009; Ballot et al. 2011; Mathur et al. 2010, 2011) open a new era in this field. In this work we presented a review on the seismic signatures of magnetic activity found in some CoRoT stars as well as the complementary observations done in Ca H\&K of the star HD~49933. 

\section{Observations and discussion}
Seismic observations provide the ability to pierce inside the stars. The eigenmodes are perturbed by the presence of magnetic fields and thus, monitoring the properties of the acoustic modes with time is a powerful tool to measure magnetic activity cycles. Moreover, depending on the modes that are affected or not by this magnetic perturbations, we can establish the depth in which this perturbations are the strongest (e.g. Salabert et al. 2011).

CoRoT provides an excellent set of long and continuous data sets of solar-like stars in which we can study the presence of any variation of the p-mode characteristics.  In particular, we are interested in fast rotating stars as they are the ones more likely to have a short activity cycles (Ossendrijver, 1997, Jouve et al. 2010). A good example could be the F2 star HD~181420 where a rotation period of $\sim$~2.6 days (Barban et al. 2009) has been measured. However, the preliminary analysis of the average frequency shift of the p modes and the maximum amplitude of the p-mode hump, $A_{\rm max}$, seems to show no clear signature of a short magnetic activity cycle. 

Another interesting star is HD~52265 (Ballot et al. 2011), a G0V metal-rich, main-sequence star. Apart from being one of the best targets measured by CoRoT so far, it is also the host of a non transiting exoplanet. The preliminary temporal analysis of the frequency shifts and  $A_{\rm max}$ reveals a small gradient in both parameters that are anticorrelated as we would expect from a star which is in the rising phase of a cycle. We are now analyzing complementary ground-based spectropolarimetric data taken by the NARVAL instrument located at the Pic du Midi Observatory. 

Finally, we have now more than a year of observations of HD~49933 in Ca H\&K lines (with only 6 months continuity because this star is eclipsed for 6 months by the Sun). Preliminary results confirm the existence of a short period magnetic activity cycle in this star as it was uncovered by CoRoT a year ago (Garc\'\i a et al. 2010). Unfortunately, the star was again eclipsed by the Sun and we could not established the length of the cycle. Incoming observations (asteroseismic and spectroscopic) during the next fall will probably help us to accurately determine the length and the properties of this magnetic cycle.

%\begin{figure}[h]
%\begin{center}
%\epsfig{width=5cm,angle=90, file=fft.eps}
%\epsfig{width=7.5cm,file=ED_minimal_list.eps}
%\caption{cap}
%\end{center}
%\end{figure}

%\vspace{3cm}
\acknowledgments{The CoRoT space mission has been developed and is operated by CNES, with contributions from Austria, Belgium, Brazil, ESA (RSSD and Science Program), Germany and Spain. RAG acknowledge the support given by the French PNPS program and CNES for the support of the CoRoT activities at the SAp, CEA/Saclay. NCAR is supported by the National Science Foundation. DS acknowledges the support from CNES.\\
}

\end{document}